\documentclass[aip,graphicx, reprint]{revtex4-2}

\draft % marks overfull lines with a black rule on the right

\usepackage{graphicx}% Include figure files
\usepackage{dcolumn}% Align table columns on decimal point
\usepackage{bm}% bold math
\usepackage[utf8]{inputenc}
\usepackage[T1]{fontenc}
\usepackage{mathptmx}
\usepackage{etoolbox}
\usepackage{color}

\makeatletter
\def\@email#1#2{%
 \endgroup
 \patchcmd{\titleblock@produce}
  {\frontmatter@RRAPformat}
  {\frontmatter@RRAPformat{\produce@RRAP{*#1\href{mailto:#2}{#2}}}\frontmatter@RRAPformat}
  {}{}
}%
\makeatother
\begin{document}

% Use the \preprint command to place your local institutional report number 
% on the title page in preprint mode.
% Multiple \preprint commands are allowed.
%\preprint{}

\title{Cathodoluminescence monitoring of quantum emitter activation in hexagonal boron nitride} %Title of paper

% repeat the \author .. \affiliation  etc. as needed
% \email, \thanks, \homepage, \altaffiliation all apply to the current author.
% Explanatory text should go in the []'s, 
% actual e-mail address or url should go in the {}'s for \email and \homepage.
% Please use the appropriate macro for the type of information

% \affiliation command applies to all authors since the last \affiliation command. 
% The \affiliation command should follow the other information.

\author{S\'ebastien Roux$^{1,2,*}$, Clarisse Fournier$^{1,*}$, Kenji Watanabe$^3$, Takashi Taniguchi$^4$, Jean-Pierre Hermier$^1$, Julien Barjon$^1$, Aymeric Delteil$^{1,\dagger}$}
%\email[]{Your e-mail address}
%\homepage[]{Your web page}
%\thanks{}
%\altaffiliation{}
\affiliation{$^1$ Universit\'e Paris-Saclay, UVSQ, CNRS,  GEMaC, 78000, Versailles, France. \\
$^2$ Université Paris-Saclay, ONERA, CNRS, Laboratoire d'étude des microstructures, 92322, Châtillon, France. \\
$^3$ Research Center for Functional Materials, 
National Institute for Materials Science, 1-1 Namiki, Tsukuba 305-0044, Japan \\
$^4$ International Center for Materials Nanoarchitectonics, 
National Institute for Materials Science, 1-1 Namiki, Tsukuba 305-0044, Japan \\
$^*$ These two authors contributed equally.\\
$\dagger$ aymeric.delteil@uvsq.fr}

% Collaboration name, if desired (requires use of superscriptaddress option in \documentclass). 
% \noaffiliation is required (may also be used with the \author command).
%\collaboration{}
%\noaffiliation

\date{\today}

\begin{abstract}
The ability to locally activate or generate quantum emitters in two-dimensional materials is of major interest for the realization of integrated quantum photonic devices. In particular, hexagonal boron nitride (hBN) has recently been shown to allow a variety of techniques for obtaining quantum emitters at desired locations. Here, we use cathodoluminescence (CL) to monitor \textit{in situ} the local activation of color centers by an electron beam in hBN. We observe that the CL signal saturates at a given surface dose, independently of the electron current density. Based on photoluminescence and photon correlations, we show that the number of photoactive color centers is proportional to the CL signal, and we estimate the maximum density of quantum emitters that can be generated by our technique. Our results provide insights about the activation mechanism and could help to optimize the controlled generation of single photon sources in hexagonal boron nitride.
\end{abstract}

% insert suggested PACS numbers in braces on next line

\maketitle %\maketitle must follow title, authors, abstract and \pacs

The recent discovery of quantum emitters in two-dimensional materials~\cite{chakraborty15, he15, koperski15, srivastava15, tonndorf15, tran16} has unveiled new roads towards integrated quantum photonics. Among the most promising 2D platforms, hBN has been shown to host a broad diversity of high-quality color centers, emitting in various wavelength ranges, from ultraviolet to near infrared~\cite{tran16, bourrelier16, martinez16, Kianinia22, Gottscholl20}. A large fraction of the current research effort is devoted towards control of the position and wavelength of these single-photon emitters (SPEs) for integration into photonic structures~\cite{ziegler19,proscia18,Xu21}. In this context, we have previously demonstrated the possibility to locally activate a recently discovered family of quantum emitters using an electron beam~\cite{Fournier21}. These blue-emitting color centers -- abbreviated B-centers in the following -- have a reproducible emission in the blue range, with very attractive optical properties, such as a high stability and a remarkably low inhomogeneous broadening of the SPE ensembles~\cite{Fournier21, Gale22, shevitski19}. However, their microscopic structure, as well as the mechanisms at the origin of their generation, remains to be understood. The current understanding is that they stem from the modification of a complex that preexists in the as-grown hBN crystal~\cite{Gale22, shevitski19}. We therefore refer to the generation process as an activation.
Here, we demonstrate the possibility to measure the cathodoluminescence (CL) of the B-centers \textit{in-situ} during the irradiation process. We subsequently characterize the sample in photoluminescence (PL) to estimate the number of optically stable SPEs generated by our technique and therefore to infer the efficiency of the method. We expect this work to provide insights on the color center activation mechanism.

 \begin{figure}
 \includegraphics[width=0.98\linewidth]{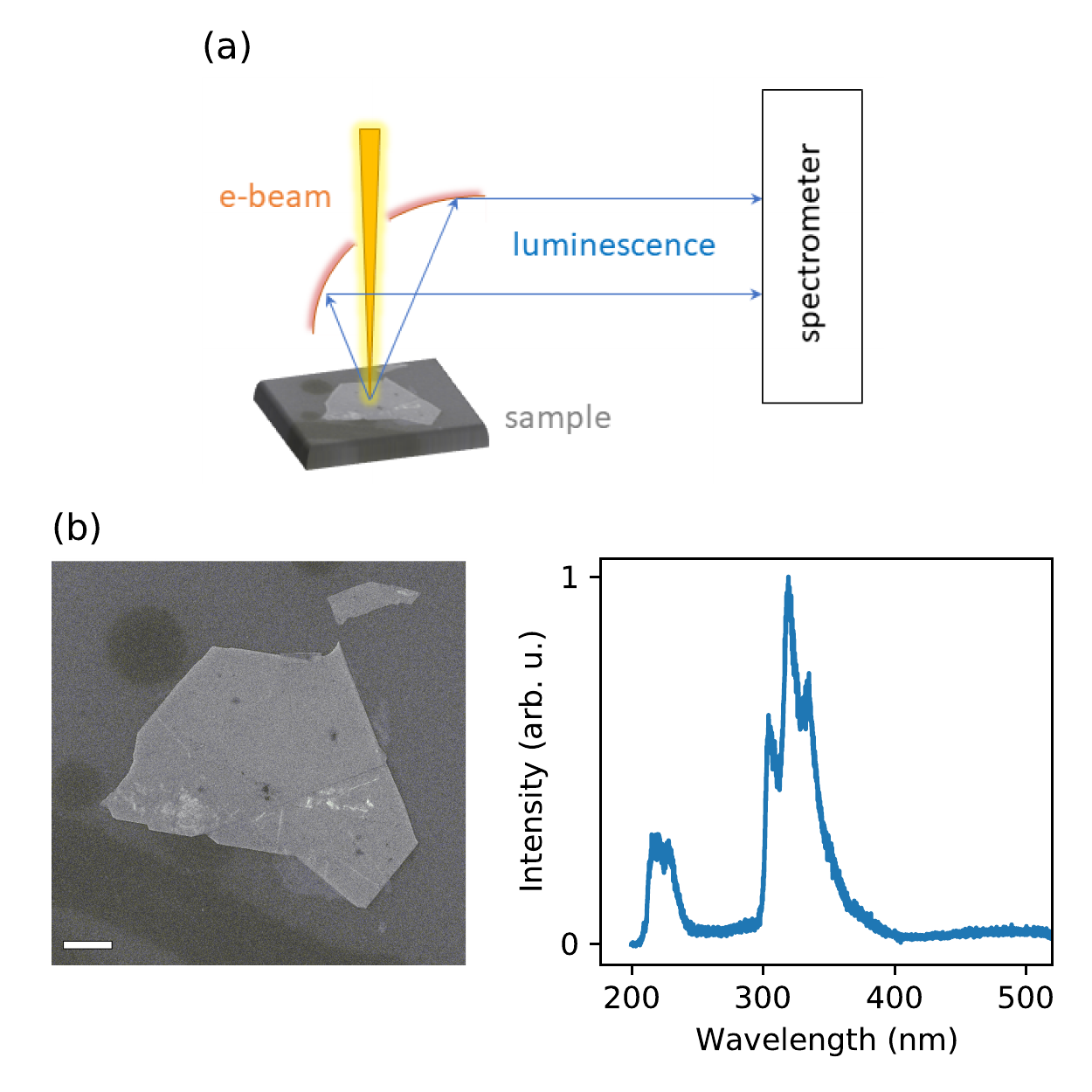}%
 \caption{\label{fig1} (a) Sketch of the experimental setup used for the electron irradiation and the CL monitoring. (b) Left: SEM image of the irradiated hBN crystal. The white scale represents 10~$\mu$m. Right: CL spectrum of the crystal prior to irradiation.}%
 \end{figure}
 
We exfoliated a hBN flake grown using the high-pressure high-temperature technique~\cite{taniguchi07} on a SiO$_2$/Si substrate. Figure~\ref{fig1}b shows a scanning electron microscope (SEM) image of the flake. Its thickness has been measured by atomic force microscopy to be 160~nm. CL is measured by collecting the sample emission by an aluminium parabolic mirror, with a hole to accommodate for the electron beam (figure~\ref{fig1}a). The bare CL spectrum of the flake is shown on figure~\ref{fig1}b, where the free exciton can be observed at 215~nm, as well as three peaks around 305~nm often attributed to emission from carbon defects~\cite{Onodera19}. The presence of this sub-bandgap emission has recently been identified as a prerequisite for the generation of B-centers in hBN~\cite{Gale22}. No PL emission from B-centers is observed prior to the irradiation process.

\begin{figure}
 \includegraphics[width=0.95\linewidth]{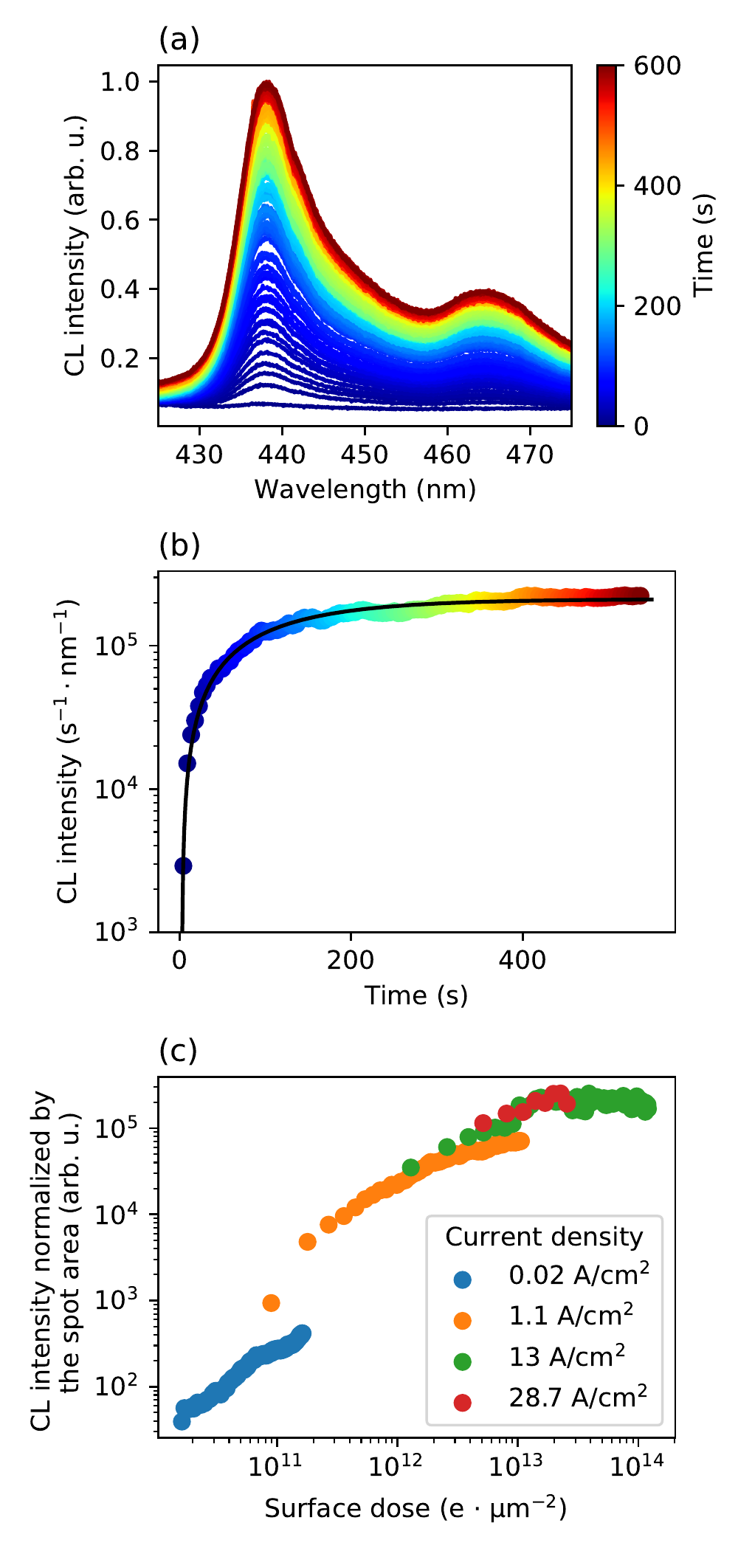}%
 \caption{\label{fig2} (a) CL spectra measured continuously with an integration time of 3~s. (b) Dots: maximum of the CL signal as a function of time. Solid line: exponential fit to the data. (c) CL signal normalized by the irradiation spot area as a function of the surface dose. Four diameters for the irradiation spot were used to vary the electron current density.}%
 \end{figure}

We have then irradiated the hBN crystal using a defocused electron beam of 1~$\mu$m diameter in a SEM. The irradiation conditions are kept constant during the whole process, with a voltage of 15~kV and a current of 9~nA. We continuously recorded the CL spectra during the irradiation, using a long-pass filter to eliminate the second order diffraction of the exciton luminescence. Figure~\ref{fig2}a shows the CL spectra taken at time intervals of 3~s. The emergence of the typical B-center emission peaks can be clearly observed. Their room-temperature spectra are composed of a zero-phonon-line at 440~nm and a red-shifted optical phonon replica $\sim$160~meV away. Figure~\ref{fig2}b shows the integrated intensity as a function of time. After a stark linear increase, the emission rapidly reaches a plateau, which we attribute to saturation of the SPE activation.

We repeated the experiment at various positions on the crystal while varying the beam diameter between 170~nm and 7.5~$\mu$m. The voltage and current are kept identical -- therefore the current density varies by more than three orders of magnitude. Figure~\ref{fig2}c shows the CL intensity as a function of the dose per unit area for four irradiation spots. Despite some slight variations between the different irradiations, it can be seen that the CL signal follows the same general trend, with a linear increase for surface doses smaller than $\sim 10^{13}$~e $\cdot \mu$m$^{-2}$ followed by a saturation around $10^{14}$~e $\cdot \mu$m$^{-2}$. The CL signal per unit area at saturation is almost independent of the beam diameter. These observations further suggest that electron-beam activation of B-centers is bounded to a maximum density of emitters.

 \begin{figure*}
 \includegraphics[width=\linewidth]{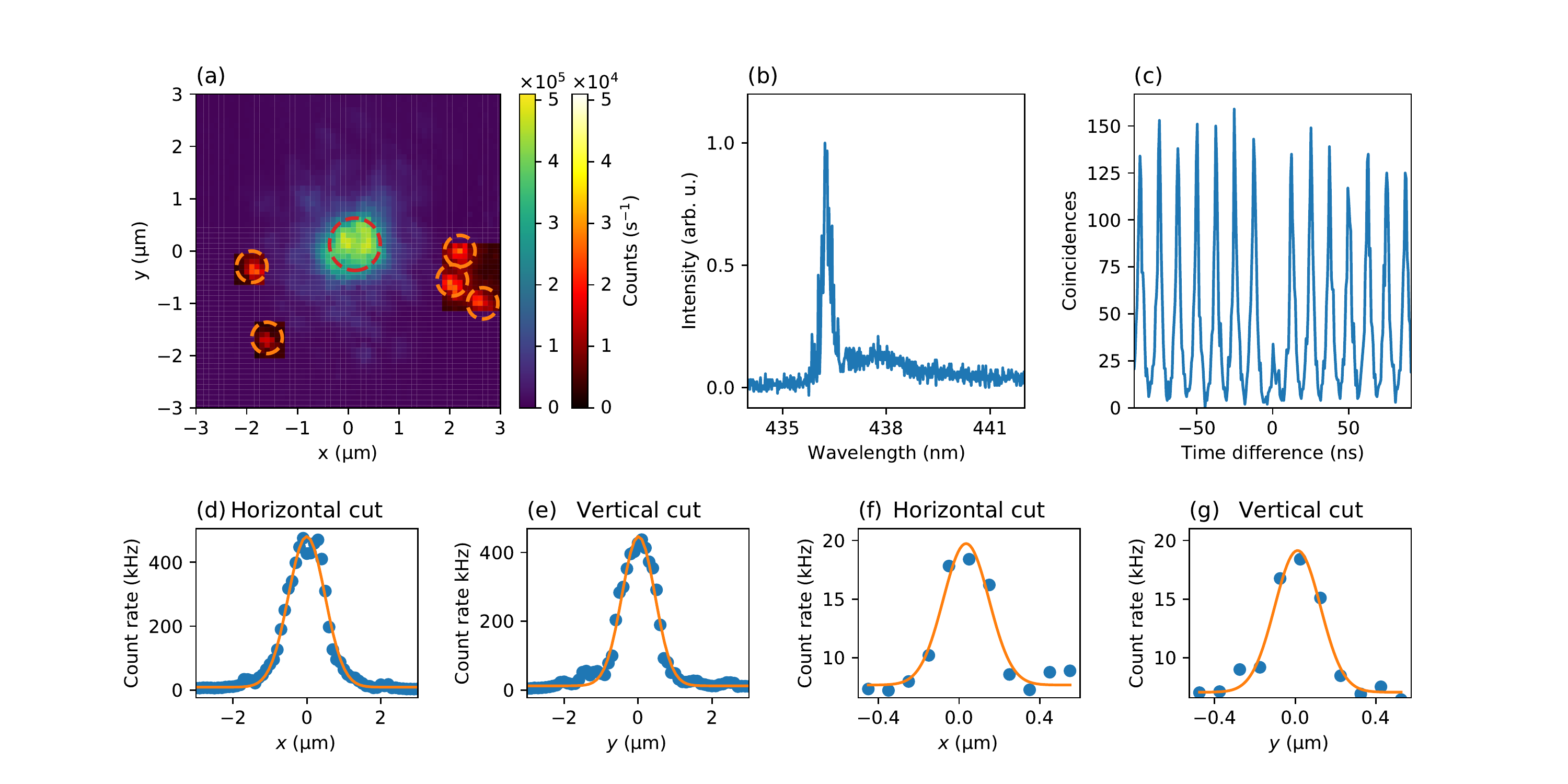}%
 \caption{\label{fig3} (a) Confocal PL map of one of the irradiated zones. The red circle denotes the position and size of the electron beam. Inside the orange circles, a second color scale with $10 \times$ smaller magnitude was used to reveal the weak luminescence of individual emitters. (b) Low temperature spectrum of the center area. (c) Second-order correlation function of one of the highlighted individual emitters, yielding $g^{(2)}(0) = 0.25 \pm 0.02$. (d), (e) Horizontal and vertical cuts of the center spot. The orange lines are Gaussian fits to the data. (f), (g) Horizontal and vertical cuts around an individual emitter. The orange lines are Gaussian fits to the data. }%
 \end{figure*}

In order to estimate the number of photostable emitters generated by the irradiations, we measured the PL of the flake in a confocal microscope, by exciting with a circularly polarized laser of wavelength 405~nm and power 400~$\mu$W. The PL signal is collected by a high-NA (0.95) air objective, filtered using a bandpass filter centered around the zero-phonon-line (ZPL) of B-centers, and conveyed to avalanche photodiodes. Figure~\ref{fig3}a shows a room-temperature PL confocal map measured around a zone irradiated by a 1~$\mu$m diameter beam, denoted by the red circle. The PL signal is the highest within the irradiation spot, and rapidly decreases with increasing distance. A horizontal and a vertical cut of the confocal map are shown on figure~\ref{fig3}d and e. These PL profiles are well fitted by Gaussian functions centered at the position of the electron beam. Accordingly, the SPE density decreases exponentially with distance from the spot center. Therefore, at a distance of 1 to 2~$\mu$m away from the center area, the density is low enough that individual SPEs can be observed. No SPE are found at distances greater than 3~$\mu$m from the spot center. We note that backaction from the substrate during irradiation could also play a role in the low but finite density of SPEs around the spot center. Further studies based on suspended hBN flakes as well as numerical simulations may help determine this possibility. The color scale of figure~\ref{fig3}a is magnified in regions where SPEs can be individually measured, as marked by the orange dashed circles. Figure~\ref{fig3}b shows a low-temperature spectrum of the center area. The characteristic ZPL at 436~nm can be observed, flanked by the acoustic phonon shoulder~\cite{Fournier21}. The narrow ensemble distribution ($< 0.3$~nm) is typical of ensembles of electron-activated B-centers. Figure~\ref{fig3}c shows the second-order correlation function of a representative individual emitter, with $g^{(2)}(0) = 0.25 \pm 0.02$, signaling a single photon emitter. All emitters highlighted with an orange circle on figure~\ref{fig3}a are individual single-photon sources exhibiting $g^{(2)}(0) < 0.5$.

In order to estimate the number of individual emitters in the whole irradiated area, we have estimated the integrated intensity associated with individual emitters~\cite{Hollenbach22}. A horizontal and a vertical cut of the confocal map around a single color center are shown figure~\ref{fig3}f and~g. These profiles are then fitted with Gaussian functions to extract the integrated PL intensity associated with the SPE. We have repeated this estimation for the five highlighted SPEs to obtain the typical integrated PL intensity associated with individual SPEs as well as its standard deviation. We have then estimated the total integrated intensity of the SPE ensemble using a similar procedure, based on the PL intensity cuts shown figure~\ref{fig3}d and~e. The ratio between the total and the individual intensities then provides an estimation of the number of emitters~\cite{Hollenbach22}, which we found to be $528 \pm 190$ for this particular spot. Knowing the spot area as well as the flake thickness, we can deduce the SPE density in the irradiated zone, which we find to be $\sim 3 \cdot 10^{15}$~cm$^{-3}$.
 \begin{figure}[hb]
 \includegraphics[width=0.8 \linewidth]{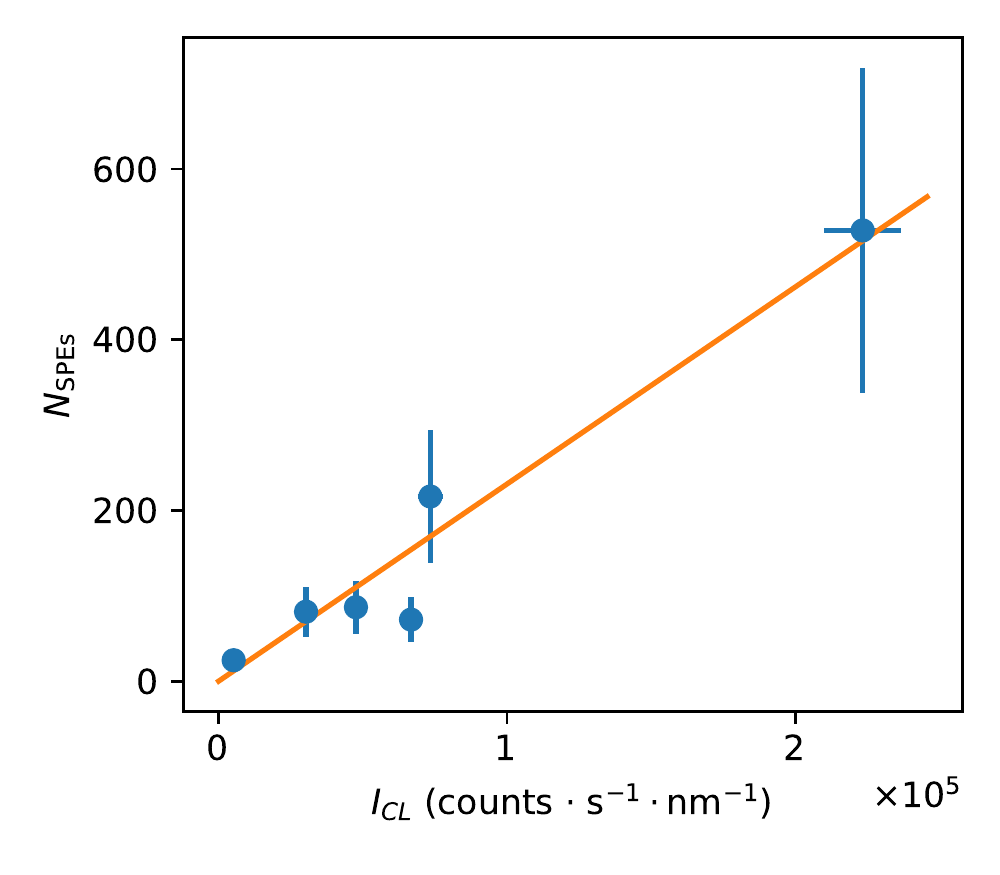}%
 \caption{\label{fig4} Blue dots: Number of SPEs estimated from PL measurements as a function of the CL signal at the end of the irradiation process. Orange line: Linear fit to the data.}%
 \end{figure}

We have repeated the process for overall six irradiation spots of different diameters on the same flake. Figure~\ref{fig4} shows the estimated number of SPEs as extracted from the PL measurements shown figure~\ref{fig3} as a function of the CL signal as measured at the end of the irradiation process. The number of SPEs is found to be proportional to the CL signal, showing that the latter constitutes an \textit{in situ} indicator of SPE activation. Through a slight improvement of our collection efficiency, it should be possible to monitor the CL signal at the scale of individual emitters. This could be used to further investigate the SPE activation process. The linear relation between the CL and PL also allows us to infer the electron dose per SPE below saturation, which we found to be in the order of $4 \cdot 10^{10}$~electrons per activated SPE at our voltage conditions (15~kV). This offers the possibility to fix \textit{a priori} the value of the irradiation dose in order to obtain a certain number of SPEs at a given position.

In summary, we have demonstrated that CL allows us to monitor the electron-beam activation of single photon emitters in hBN. Our measurements have revealed a saturation of the activation process, which could help to understand the physical origin of the color centers as well as the irradiation mechanism. The fact that the SPE number saturates at a relatively low density of $3 \cdot 10^{15}$~cm$^{-3}$ is compatible with the current understanding that the irradiation modifies a pre-existing specie of finite density, rather than creating an intrinsic defect from pristine crystal~\cite{Fournier21, Gale22, shevitski19}. This limit density corresponds to a mean distance of 70~nm between neighboring centers. This lengthscale gives the typical spatial accuracy with which a single color center can be activated. While it may not be sufficient for some near-field optics applications such as controlled coupling to a plasmonic resonator, it is on the other hand compatible with deterministic integration into dielectric photonic structures.

% Body of paper goes here. Use proper sectioning commands. 
% References should be done using the \cite, \ref, and \label commands
%\section{}
%%\label{}
%\subsection{}
%\subsubsection{}

% If in two-column mode, this environment will change to single-column format so that long equations can be displayed. 
% Use only when necessary.
%\begin{widetext}
%$$\mbox{put long equation here}$$
%\end{widetext}

% Figures should be put into the text as floats. 
% Use the graphics or graphicx packages (distributed with LaTeX2e).
% See the LaTeX Graphics Companion by Michel Goosens, Sebastian Rahtz, and Frank Mittelbach for examples. 
%
% Here is an example of the general form of a figure:
% Fill in the caption in the braces of the \caption{} command. 
% Put the label that you will use with \ref{} command in the braces of the \label{} command.
%

% Tables may be be put in the text as floats.
% Here is an example of the general form of a table:
% Fill in the caption in the braces of the \caption{} command. Put the label
% that you will use with \ref{} command in the braces of the \label{} command.
% Insert the column specifiers (l, r, c, d, etc.) in the empty braces of the
% \begin{tabular}{} command.
%
% \begin{table}
% \caption{\label{} }
% \begin{tabular}{}
% \end{tabular}
% \end{table}
\section*{Supplementary Material}
See supplementary material for a description of the characterization of the electron beam diameter and additional PL confocal maps.

% If you have acknowledgments, this puts in the proper section head.
\begin{acknowledgments}
 The authors acknowledge Aur\'elie Pierret and Michael Rosticher for the flake exfoliation. This work is supported by the French Agence Nationale de la Recherche (ANR) under reference ANR-21-CE47-0004-01 (E$-$SCAPE project). This work also received funding from the European Union’s Horizon 2020 research and innovation program under Grant No. 881603 (Graphene Flagship Core 3). K.W. and T.T. acknowledge support from JSPS KAKENHI (Grant Numbers 19H05790, 20H00354 and 21H05233).
\end{acknowledgments}

% Create the reference section using BibTeX:
%\bibliography{your-bib-file}
~\\
%
%
%\section*{Data Availability Statement}
%The data that support the findings of this study are openly available in
%[repository name] at http://doi.org/[doi], reference number [reference number].

\end{document}